\documentclass[floatfix,showkeys,pra,showpacs,preprint,amsmath,superscriptaddress]{revtex4-1}
\usepackage{hyperref}
\usepackage[utf8]{inputenc}
\usepackage{graphicx}
\usepackage{color}
\usepackage{bbding}
\usepackage{amssymb}
\usepackage{booktabs}

\begin{document}

\title{Experimental realization of double Bragg diffraction: robust beamsplitters, mirrors, and interferometers for Bose-Einstein condensates}
\author{J. K\"uber}
\author{F. Schmaltz}
\author{G. Birkl}
\affiliation{Institut für Angewandte Physik, Technische Universität Darmstadt, Schlossgartenstraße 7, D-64289 Darmstadt}

\date{\today}
\begin{abstract}
We present the experimental implementation of double Bragg diffraction of Bose-Einstein condensates (BECs) as proposed in [E. Giese, A. Roura, G. Tackmann, E. M. Rasel, and W. P. Schleich, Phys. Rev. A \textbf{88}, 053608 (2013)]. We excite Rabi oscillations between the three coupled momentum states $\left| 0 \hbar k \right\rangle$ and  $\left| \pm 2 \hbar k \right\rangle$. By selecting appropriate interaction times we generate highly efficient beamsplitters and mirrors for Bose-Einstein condensates. In addition, we demonstrate higher-order double Bragg diffraction and display beamsplitters with up to $ \pm 6 \hbar k$ momentum transfer. We compare double Bragg diffraction to several other experimental realizations of beamsplitters. Finally, we show that double Bragg diffraction is well suited for matter wave interferometry by realizing a Ramsey-type interferometer in a quasi one-dimensional waveguide.
\end{abstract}
\maketitle
\begin{figure}
	\centering
		\includegraphics[width=0.49\textwidth]{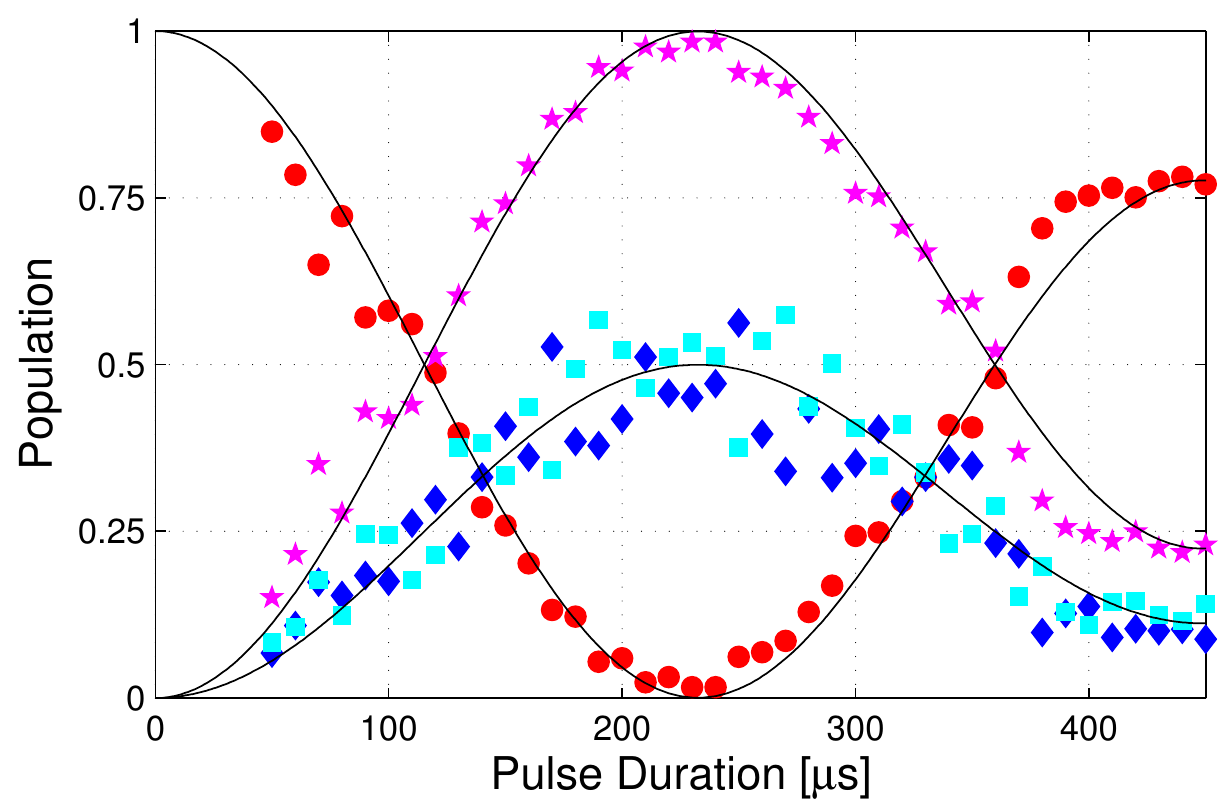}
	\caption{Rabi oscillations between the momentum states $\left| 0 \hbar k \right\rangle$ ($\circ$),
$\left| + 2 \hbar k \right\rangle$ ($\lozenge$), and $\left| - 2 \hbar k \right\rangle$ ($\square$) of a $^{87}$Rb BEC in a double Bragg lattice. The atoms oscillate with $\Omega_{eff} = 2 \pi \cdot \left(1.09 \pm 0.06 \right)\,$kHz. For a $230\,\mu$s pulse, a robust beamsplitter is realized with an efficiency of $99\,$\% for the combined population of the momentum states $\left| \pm 2 \hbar k \right\rangle$ (\FiveStarOpen) and an almost complete depletion of state $\left| 0 \hbar k \right\rangle$ ($\circ$). The solid lines are given by Eq. (\ref{form:solution1}) with no free parameters. 
}
	\label{fig:plot_2hk_good_contrast}
\end{figure}
\begin{figure}
	\centering
		\includegraphics[width=0.45\textwidth]{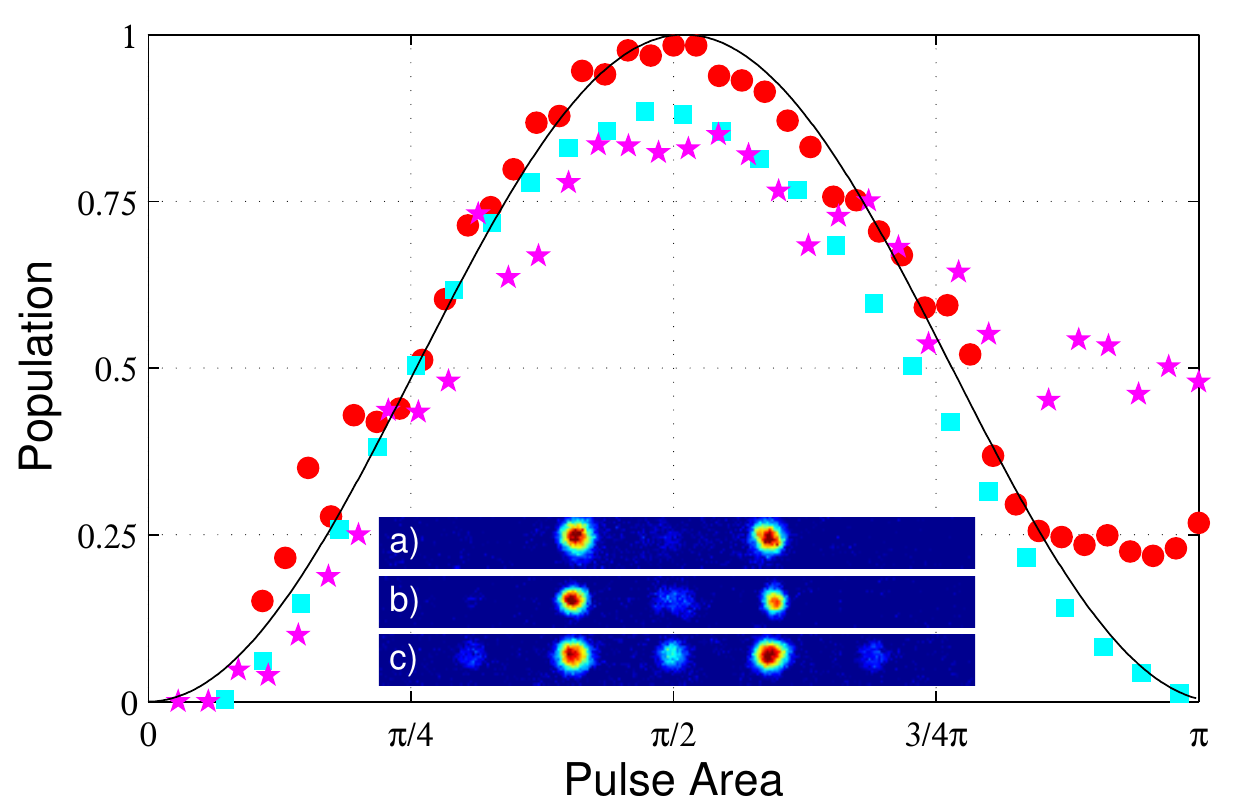}
	\caption{Comparison of three implementations of beamsplitters for $\left| \pm 2 \hbar k \right\rangle$: The fraction of atoms in momentum states $\left| \pm 2 \hbar k \right\rangle$ after a lattice pulse of variable duration $\tau$ is depicted for a pulse area of up to $\pi$. Double Bragg diffraction ($\circ$, inset (a)) shows that $99\,$\% of the atoms can be transferred into $\left| \pm 2 \hbar k \right\rangle$. By using three-frequency Bragg diffraction (\FiveStarOpen, inset (b)) we achieve an efficiency of $82\,$\%. The 'Pendellösung' ($\square$, inset (c)) yields a maximum transfer efficiency of $88\,$\%. 
The solid line indicates a perfect Rabi oscillation.}
	\label{fig:pm2hkcomparisson}
\end{figure}
\begin{figure}
	\centering
		\includegraphics[width=0.45\textwidth]{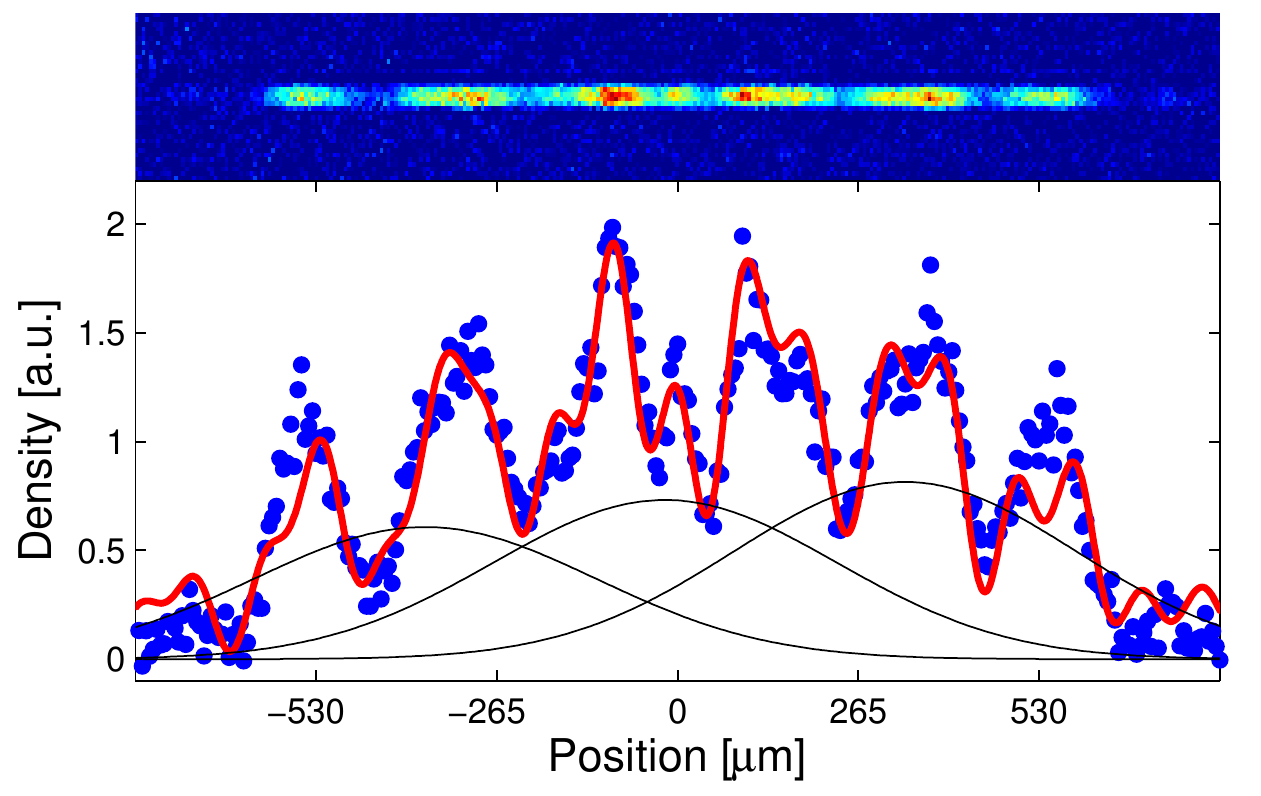}
	\caption{Double Bragg interferometer: Density distribution after a Ramsey-type interferometer sequence of two $\pi/2$ pulses with a pulse separation of $T = 105\,\mu$s. The interferometer yields a density distribution consisting of three output ports (depicted as black Gaussian profiles) modulated by two interferometer patterns with different spatial periods.}
	\label{fig:interference_period}
\end{figure}

The application of atom-optical and atom-interferometrical techniques \cite{RevModPhys.81.1051} to ultra-cold atoms and Bose-Einstein condensates is in the stage of revolutionizing high-precision measurements and time-keeping. In this context, Bragg diffraction in standing or moving optical lattices has been shown to be a versatile tool in matter wave optics \cite{kozuma99} and atom interferometry \cite{denschlag2000, sagnac, interferometerKetterle}, and to be an elegant way to probe BECs for a fundamental understanding of their properties \cite{braggSpectroscopy}. Bragg lattices are used as beam splitters and mirrors and are advantageous due to their scalability, simplicity, and robust experimental implementation. In this article we demonstrate the first implementation of double Bragg diffraction \cite{double_bragg} for coherent and symmetric splitting and recombination of ultracold atomic matter waves and BECs. Similar to standard Bragg diffraction, double Bragg diffraction can accelerate atoms to multiples of $2\hbar k$ momenta. In contrast, the experimental implementation of double Bragg diffraction is achieved with a single input beam path yielding a robust system that is also suited for portable setups as for micro-gravity experiments \cite{fallturm}.\\
%
\begin{figure}
	\centering
		\includegraphics[width=0.49\textwidth]{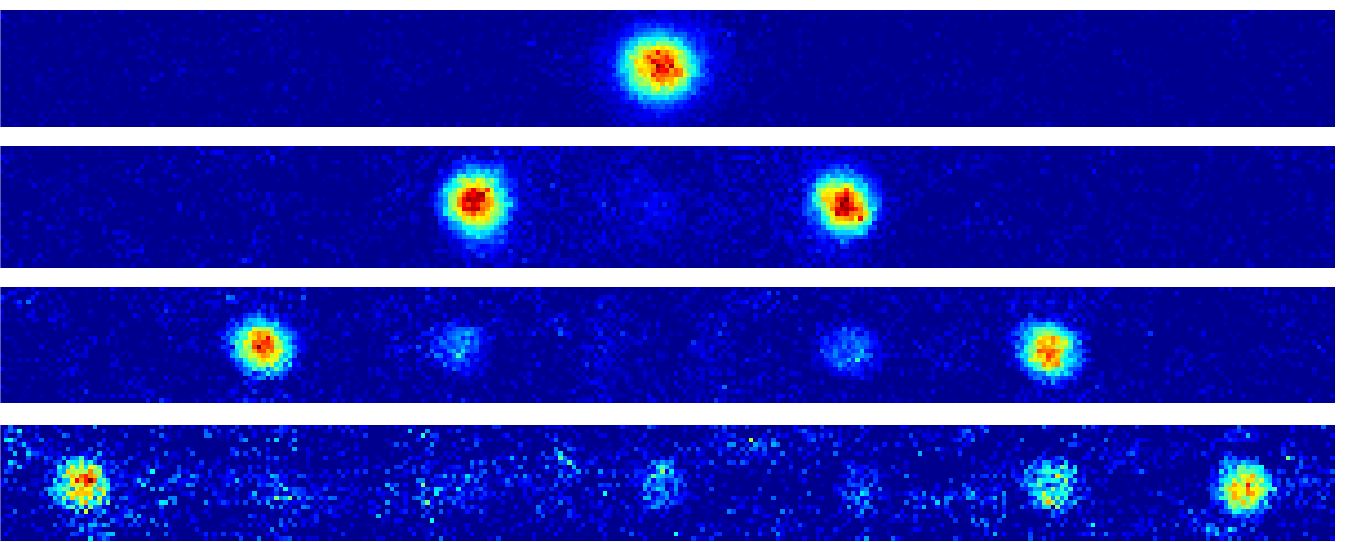}
	\caption{Demonstration of a beamsplitter for BECs of $^{87}$Rb for different orders of double Bragg diffraction. After a $\pi/2$ pulse a waiting time of $18\,$ms is applied in which the atoms move freely before absorption imaging. (1st row) BEC at rest after evaporation. (2nd row) beamsplitter for $\left| \pm 2 \hbar k \right\rangle$ with a maximum combined efficiency of $99\,$\%. (3rd row) 2nd order beamsplitter with maximum efficiency of $77\,$\% for both momentum states $\left| \pm 4 \hbar k \right\rangle$. The remainder of the atoms are excited to $\left| \pm 2 \hbar k \right\rangle$. (4th row) 3rd order beamsplitter: $74\,$\% of the atoms are transferred into the momentum state $\left| \pm 6 \hbar k \right\rangle$. Also momentum states $\left| 0 \hbar k \right\rangle$, $\left| \pm 2 \hbar k \right\rangle$, and $\left| \pm 4 \hbar k \right\rangle$ are excited.}
	\label{fig:waterfall}
\end{figure}
To observe Bragg diffraction in a moving optical lattice one has to ensure energy and momentum conservation, thus for an anti-parallel beam configuration one has to fulfill the condition \cite{kozuma99}:
\begin{equation}
  \hbar \cdot \Delta \omega = n \cdot 2\frac{\hbar^2 k^2}{m}\, , 
	\label{form:energyandmomentum}
\end{equation}
where k denotes the wave number of the Bragg beams, $\Delta \omega$ the frequency offset between the two beams, and n equals the order of Bragg diffraction. For $^{87}$Rb with mass m and recoil frequency $\omega_R = \frac{\hbar k^2}{2m}$, we find a solution for $\Delta \omega =n \cdot  4 \cdot \omega_R = n \cdot 2 \pi \cdot 15.08\,$kHz for a Bragg lattice at $780\,$nm. First-order Bragg diffraction ($n=1$) results in wave packets with momenta $\pm 2\hbar k$. To obtain higher orders of Bragg diffraction the detuning $\Delta \omega$ has to be chosen appropriately (Fig. \ref{fig:waterfall}).\\
\begin{figure}
	\centering
		\includegraphics[width=0.45\textwidth]{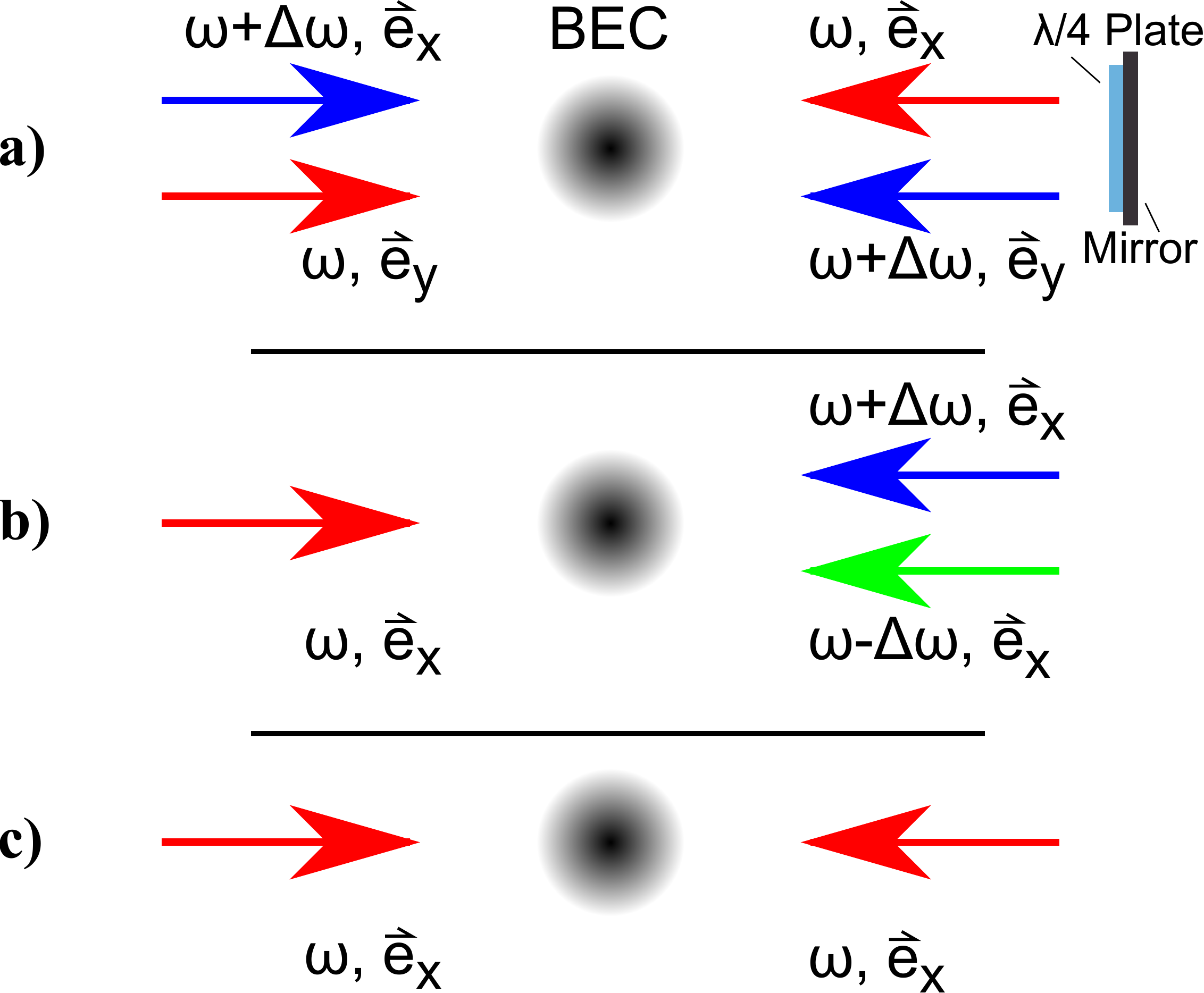}
	\caption{Different beam configurations for Bragg diffraction: a) Double Bragg diffraction uses an incoming pair of orthgonally (in our case linearly) polarized light beams with frequency offset $\Delta \omega$. A quarter wave plate and a mirror retro-reflect the beams \cite{double_bragg}. This creates a pair of one-dimensional moving optical lattices with orthogonal polarization. b) Three-frequency Bragg diffraction uses one beam with the center frequency $\omega$ and two counterpropagating beams with the same polarization but different frequencies $\omega \pm \Delta \omega$, respectively. c) The 'Pendell\"osung' configuration utilizes a stationary optical lattice.}
	\label{fig:scheme}
\end{figure}
A complete theoretical description of double Bragg diffraction can be found in \cite{double_bragg} and therefore we limit our discussion to the basic aspects. Double Bragg diffraction describes the coupling of three momentum states via four light fields. Two incoming light beams with orthogonal polarization and frequency difference $\Delta \omega$ are retro-reflected by a combination of a $\lambda/4$-waveplate and a mirror (see Fig. \ref{fig:scheme} a)). This results in two moving optical lattices with orthogonal polarizations. The configuration is viable for linear polarization as well as circular polarization for the input beams. The state dynamics can be solved analytically by the "method of averaging" \cite{double_bragg, 1961amtn.book.....B}. The solution yields an oscillating three-level system of momentum states $\left| 0 \hbar k \right\rangle$, $\left| 2 \hbar k \right\rangle$, and  $\left| - 2 \hbar k \right\rangle$ that can be described via
\begin{align}
	\begin{split}
	  \gamma\left(t\right) = {}& \left[\frac{1}{\Omega_{eff}^2} \left( \begin{matrix} \Omega^2 & -\omega_D\Omega & -\Omega^2 \\ -\omega_D\Omega & \omega_D^2 & \omega_D\Omega \\ -\Omega^2 & \omega_D\Omega & \Omega^2  \end{matrix} \right) \right. \\ \nonumber
		& + \frac{\cos\left(\Omega_{eff} t \right)}{\Omega_{eff}^2} \left( \begin{matrix} \omega_D^2 + \Omega^2 & \omega_D\Omega & \Omega^2 \\ \omega_D\Omega & 2\Omega^2 & -\omega_D\Omega \\ \Omega^2 & -\omega_D\Omega & \omega_D^2 + \Omega^2 \end{matrix} \right)  \\ \nonumber
		& \left. + \frac{i \sin\left(\Omega_{eff} t \right)}{\Omega_{eff}} \left( \begin{matrix} \omega_D & \Omega & 0 \\ \Omega & 0 & \Omega \\ 0 & \Omega & -\omega_D  \end{matrix} \right) \right] \gamma\left( 0 \right) \, ,
  \end{split}\\
	\label{form:solution1}
\end{align}
where $\gamma\left(t\right)$ represents the state vector of the system \cite{double_bragg}. $\gamma\left(0\right) = \left(0, 1, 0 \right)^{T}$ represents the initial state where all atoms are in state $\left| 0 \hbar k \right\rangle$, i.e. at rest. To make sure that higher-orders of Bragg diffraction are suppressed, the Rabi frequency $\Omega$, giving the rate of population transfer, has to be small compared to the recoil frequency $\omega_R$ \cite{double_bragg}. The momentum distribution of an atomic cloud yields a distinct Doppler shift $\omega_D$ for each component of the cloud according to its momentum. This effect is taken into account by the effective Rabi frequency $\Omega_{eff} = \sqrt{2\Omega^2 + \omega_D^2}$. The resulting superposition of oscillations damps the oscillation in the ensemble average and can result in a reduction of the overall efficiency.\\
%
%
%
%
We generate a pair of orthogonally linear polarized light fields to create a double Bragg lattice (Fig. \ref{fig:scheme} a)). The light is detuned $750\,$MHz to the blue with respect to the $F = 1 \rightarrow F = 2$ (repump) transition of the $D_2$ line at $780\,$nm. To ensure frequency stability the light is offset-locked to a stabilized reference laser. Additionally we impose a frequency offset $\Delta \omega$ between the two input fields by two accusto-optical modulators (AOMs), combine the fields with a polarizing beamsplitter and guide the light through a single optical fiber to the experiment. To create well defined lattice pulses we use a pair of arbitrary waveform generators to apply Gaussian amplitude envelopes in both AOMs. Our experimental setup uses a crossed optical dipole trap to create an all-optical BEC of 25000 $^{87}Rb$ atoms with a condensate fraction $N_C / N \geq 0.8$ and a temperature of $27\,$nK \cite{atomics_bec}. The Bragg beams are oriented collinear along one of the dipole trap legs and have a waist of $1.7\,$mm at the location of the atoms. With a laser power of about $0.25\,$mW per beam we achieve a lattice depth of $16\cdot10^{-3}\,E_R$ known within a relative uncertainty of $5\,$\%.
%
\begin{figure*}
	\centering
		\includegraphics[width=0.99\textwidth]{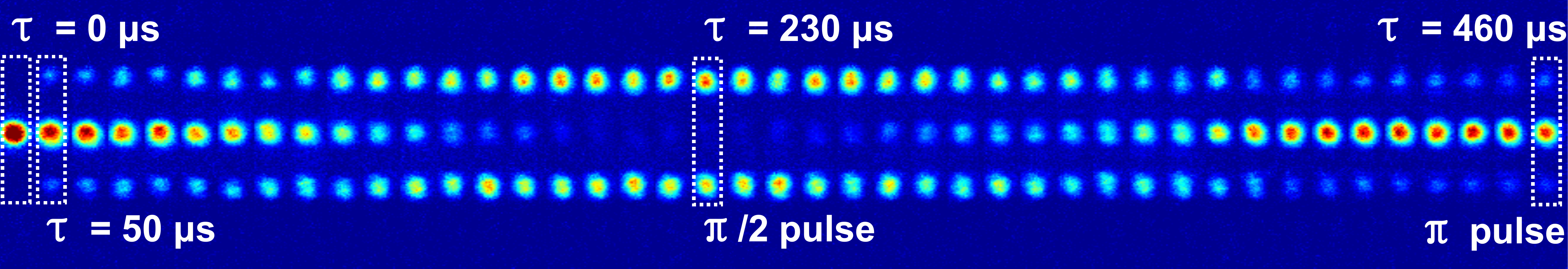}
	\caption{Density profiles of double Bragg diffracted BECs. Each column shows the density distribution after a double Bragg pulse of duration $\tau$ and additional waiting time of $18\,$ms. The atoms move according to their momentum in free space. The center row shows atoms in state $\left| 0 \hbar k \right\rangle$. At the top are atoms in state $\left| - 2 \hbar k \right\rangle$ and at the bottom in momentum state $\left| + 2 \hbar k \right\rangle$, respectively. The farthermost left column shows atoms at rest directly after evaporation without applying a Bragg pulse. The second column shows atoms after a lattice pulse of $\tau = 50\,\mu$s followed by columns with an increment of $10\,\mu$s for each column. 
}
	\label{fig:2hk}
\end{figure*}
For double Bragg diffraction we produce a BEC and immediately after turning off the dipole trap beams we apply a double Bragg pulse of variable duration $\tau$ (Fig. \ref{fig:2hk}). After that, we wait an additional $18\,$ms to let the momentum states separate before absorption imaging. The population of each momentum component is determined by fitting the density distribution at each component separately. All experimental images and the extracted data points in this article are the average of two individual experimental runs.\\
Fig. \ref{fig:2hk} shows the time evolution of double Bragg diffracted BECs in free space. Before the application of the light pulse all atoms are at rest.  Light pulses of duration $\tau$ are applied for $\tau$ between $50\,\mu$s and $460\,\mu$s immediately after the release of the BEC from the crossed dipole trap. As the pulse duration increases more atoms oscillate into the momentum states $\left| + 2 \hbar k \right\rangle$ and $\left| - 2 \hbar k \right\rangle$. For a pulse duration of $230\,\mu$s $99\,$\% of the atoms are transferred to the states $\left| \pm 2 \hbar k \right\rangle$ thus demonstrating a highly efficient beamsplitter. Fig. \ref{fig:plot_2hk_good_contrast} presents the population of each momentum state for the measurements of Fig. \ref{fig:2hk}. The Rabi oscillations between different momentum states are clearly visible. We also observe a slight variation of the relative population between the two momentum states $\left| + 2 \hbar k \right\rangle$ and $\left| - 2 \hbar k \right\rangle$ and attribute this to relative intensity or phase fluctuations between the lattice beams. Fig. \ref{fig:plot_2hk_good_contrast} includes the solution of Eq. (\ref{form:solution1}) (solid lines) for our experimental parameters with a calculated Rabi frequency of $\Omega_{eff} = 2 \pi \cdot \left(1.09 \pm 0.06 \right)\,$kHz \cite{latticedepth} based on our potential depth of $16\cdot10^{-3}\,E_R$ and assuming $w_D = 0$. The experimental data matches this theoretical prediction to a high degree. An independent measurement of the momentum width of our BEC is used to estimate a weighted mean Doppler shift $\omega_D$. This results in a Rabi frequency about 5\% larger than the one without including $w_D$. This is also observable in Fig. \ref{fig:plot_2hk_good_contrast} as a slightly faster oscillation of the experimental data. After a nearly complete momentum transfer from $\left| 0 \hbar k \right\rangle$ to $\left| \pm 2 \hbar k \right\rangle$ for $\tau = 230\,\mu$s we find that a $\pi$ pulse ($\tau = 460\,\mu$s) leaves about  $22\,$\% of the atoms in states $\left| \pm 2 \hbar k \right\rangle$. This effect is fully covered by Eq. (\ref{form:solution1}) as shown by the solid lines in Fig. \ref{fig:plot_2hk_good_contrast} for large $\tau$ and is caused by the finite momemtum width \cite{double_bragg, momentumWidth}. We could show, that for a pulse area of 2$\pi$, $98\,$\% of the atoms are transferred back to state $\left| 0 \hbar k \right\rangle$, giving an almost perfect mirror \cite{kueberthesis}.\\
We also investigated higher-order Bragg diffraction. Fig. \ref{fig:waterfall} shows the experimental results obtained for double Bragg diffraction of order $n = 1, 2, 3$. The 2nd row shows a beamsplitter with $n = 1$ and corresponds to the image at $\tau = 230\,\mu s$ in Fig. \ref{fig:2hk}. A splitting with a maximum order of $n=2$ is depicted in the 3rd row. The relative frequency shift between the two lattice beams is $\Delta\omega = 2\pi \cdot 30.16\,$kHz. We achieve a maximum combined efficiency of $77\,$\% in the momentum states $\left| + 4 \hbar k \right\rangle$ and $\left| - 4 \hbar k \right\rangle$. Due to the limited power of our laser setup, we stabilized the lattice light closer to the $^{87}$Rb repump transition. We achieve a $\pi/2$ pulse and therefore maximum splitting of the BEC in the $n=2$ momentum states after $175\,\mu$s. The shorter time is consistent with an increased oscillation frequency due to the reduced detuning.
We also were able to implement a beam splitter with $n=3$. The maximum efficiency is $74\,$\% for the momentum states $\left| \pm 6 \hbar k \right\rangle$ (Fig. \ref{fig:waterfall} (4th row)). We observe non negligible amounts of atoms in states $\left| \pm 2 \hbar k \right\rangle$ for $n=2$ and in states
$\left| 0 \hbar k \right\rangle$, $\left| \pm 2 \hbar k \right\rangle$, and $\left| \pm 4 \hbar k \right\rangle$ for $n=3$, reducing the efficiency of the beamsplitter. This is mostly caused by spontaneous scattering due to reduced detuning of the lattice beams for achieving high coupling strengths.\\
%
Next, we compare double Bragg diffraction to two other methods of beam splitting. For increased flexibility in splitting but also allowing single-momentum acceleration and deceleration we implemented three-frequency Bragg diffraction \cite{thomasring} as depicted in Fig. \ref{fig:scheme} b). For beam splitting, one beam with the center frequency $\omega$ is combined with two counterpropagating beams with the same polarization but different frequencies $\omega \pm \Delta \omega$, respectively. We choose the same lattice depth as for the case of double Bragg diffraction. As Fig. \ref{fig:pm2hkcomparisson} depicts we are able to transfer $82$\% of the initially resting atoms to the momentum states $\left| \pm 2 \hbar k \right\rangle$. The increased flexibility requiring two different optical paths reduces the efficiency of the splitting process.\\
%
We also implemented the so called 'Pendell\"osung' \cite{pendelloesung, denschlag2000} to achieve splitting of a BEC in a non-moving lattice, e.g. a lattice with no relative detuning ($\Delta \omega = 0$). For this work, we use the three-frequency beam splitter setup with counterpropagating beams having the same frequency $\omega$. A simplified arrangement can be based on a single incoming beam being retro-reflected by a standard mirror, of course. The setup now creates a single standing optical lattice (Fig. \ref{fig:scheme} c)). We chose the lattice depth to be $19\,E_R$ and observe Rabi oscillations with a $\pi/2$ time of about $9\,\mu$s. The higher lattice depth was necessary to achieve the best possible splitting. In a standing optical lattice, the oscillation frequency and the highest excited momentum state cannot be chosen independently. Fig. \ref{fig:pm2hkcomparisson} shows that we transfer $88\,$\% of the atoms into momentum states $\left| \pm 2 \hbar k \right\rangle$. The main reason for not reaching a full momentum transfer is excitation of multiple diffraction orders \cite{maxnstanding} as seen in the inset of  Fig. \ref{fig:pm2hkcomparisson}c).\\
All three methods show that splitting of the BEC is possible with an efficiency of $80\,$\% or higher. Double Bragg diffraction offers a stable setup that can be controlled in a simple way and offers the highest efficiency. Three-frequency Bragg diffraction offers the flexibility to create asymmetric splitting by imposing different values of $\Delta \omega$ to the lattice beams, but at the cost of increased complexity and reduced efficiency. A non-moving optical lattice can be made with a simple setup but has limited control of the parameters and requires a significantly increased Rabi frequency.\\
%
In order to explore the applicability of double Bragg diffraction for interferometric measurements we implemented a Ramsey-type interferometer in a quasi one-dimensional waveguide \cite{denschlag2000, thomasring}. After preparation of the BEC we increase the intensity of the dipole trap leg collinear with the Bragg lattice sufficiently to hold the atoms against gravity and simultaneously turn off the other leg. We perform an interferometry sequence of two $\pi/2$ pulses separated by a variable time $T$. To ensure that the mean-field energy is depleted completely and ballistic expansion is dominant \cite{castin96, bectimeevolution} we introduce a waiting time of $30\,$ms before the first $\pi/2$ pulse. Depending on the pulse separation $T$, we create a variable spatial separation $\Delta x(\Delta p)$ of the atomic wavepackets produced by the first $\pi/2$ beamplitter. The separation $\Delta x(\Delta p)$ gives a direct control of the fringe period $d$ created interferometrically by applying a second $\pi/2$ pulse. The displaced wave functions get projected onto the same momentum states which results in periodic density modulations due to their phase difference with the period $d$ being inversely proportional to $\Delta x(\Delta p)$ \cite{castin96, denschlag2000}. After the second $\pi/2$-pulse an additional waiting time of $30\,$ms in the waveguide separates the resulting momentum states before detection.
%
%
%
%
Fig. \ref{fig:interference_period} shows an interference pattern generated by the described Ramsey interferometer. The resulting density distribution shows a combination of three atom clouds representing the three output ports with momenta $0\hbar k$ and $\pm 2 \hbar k$. Each output features a density modulation with two discrete periods separated by a factor of 2. The smaller period can be attributed to a relative momentum $\Delta p = 4 \hbar k$ of the partial wavefunctions traveling with respective momenta of $\pm 2\hbar k$ in opposite directions. Additionally, the interferometer shows a second period that can be attributed to a relative momentum of $\Delta p = 2\hbar k$. This structure is explained by non-perfect beamsplitting, e.g. caused by the larger momentum spread in the waveguide configuration, leading to the additional occurrence of atoms with momentum $0\hbar k$ within the interferometer. The solid red line in Fig. \ref{fig:interference_period} depicts a fit to the density distributions consisting of a total of three output ports with the two spatial density modulations. Each fitted period matches the calculated fringe period for the respective displacement $\Delta x(\Delta p)$. Underlying are three Gaussian distributions indicating the envelopes of the output ports after the interferometer sequence.\\
In summary, we have described the first experimental realization of double Bragg diffraction with a one-dimensional optical lattice. Our experimental results follow the theoretical description first given in \cite{double_bragg}. We showed that excitation of first and higher momentum orders is possible with double Bragg diffraction. Additionally, we compared the coherent splitting of double Bragg diffraction with two other methods for the creation of matter wave beamsplitters. A first implementation of double Bragg diffraction interferometry was demonstrated: We used double Bragg diffraction to image the phase profile of a Bose-Einstein condensate by applying a Ramsey-type pulse sequence. These measurements confirm that double Bragg diffraction is an important novel tool for interferometric measurements with ultracold atoms.\\
\begin{acknowledgments}
We would like to thank Reinhold Walser, Mathias Schneider, and Luis Fernando Barragán Gil for helpful discussions. During the preparation of this manuscript we learned that related experiments are performed by the group of E. Rasel and W. Ertmer (Leibniz Universit\"at Hannover) in collaboration with the group of W. Schleich (Universit\"at Ulm).
\end{acknowledgments}
\bibliographystyle{apsrev4-1}
\bibliography{double_bragg}
\end{document}